\documentclass[epjCONF]{svjour}
\usepackage{amsmath}
\usepackage{graphics}
\usepackage[varg]{txfonts} 
\usepackage[latin1]{inputenc}
\usepackage[final]{graphicx}

\session-title{Hot and Cold Baryonic Matter -- HCBM 2010}
\begin{document}
\title{Underlying events in $p+p$ collisions at LHC energies}
\author{Andr\'as G. Ag\'ocs\inst{1,2}\fnmsep\thanks{\email{agocs@rmki.kfki.hu}} 
\and Gergely G. Barnaf\"oldi\inst{1} \and P\'eter L\'evai\inst{1} }
\institute{KFKI Research Institute for Particle and Nuclear Physics of the HAS, \\
  29-33 Konkoly-Thege M. Str. H-1121 Budapest, Hungary \and  E\"otv\"os University, \\
  1/A P\'azm\'any P\'eter S\'et\'any, H-1117 Budapest, Hungary}
%
\abstract{
General properties of hadron production are investigated 
in proton-proton collisions at LHC energies.
We are interested in the characteristics 
of hadron production outside the identified jet cones. We improve
earlier definitions and introduce surrounding rings/belts 
around the cone of identified jets. In this way even multiple
jet events can be studied in details. We define the underlying event as
collected hadrons from outside jet cones and outside surrounding belts,
and investigate the features of these hadrons.
We use a PYTHIA generated data sample of proton-proton collisions at $\sqrt{s}=7$ TeV.
This data sample is analysed by our new method and 
the widely applied CDF method.
Angular correlations and momentum distributions  have been
studied and the obtained results are compared and discussed. }
%
\maketitle
%
%

\section{Introduction}

The production of hadron showers with large transverse energy (named as "jets")
is one of the most interesting phenomena in high energy proton-proton 
(and hadron-hadron) collisions. Jet production is explained by the interaction
of the ingredients of protons, namely partons: quarks, antiquarks, and gluons.
The perturbativ quantum chromodynamics (pQCD) improved parton model~\cite{FieldQCD}
can describe the momentum distribution of produced
secondary partons very successfully. The hadronization of the secondary partons
is followed by parton fragmentation functions and the resulted hadron showers
carry the energy and momenta of the original partons.
Jet analysis efforts are aiming the identification of these original partons
and to determine their properties.
However, during proton-proton ($pp$) collisions the leading jet production is only
one part of the activity, in parallel many partonic subprocesses take place with 
smaller interaction energy.  

First of all, multiple parton-parton collisions
with relatively large momentum exchange can happen with a reasonable probability
generating the problem of multiple-jet identification. At LHC energies these
multiple-jet events have a reasonable yield, their investigation will become one
of the focus study of strong interaction in forthcoming LHC experiments.
These interactions form a background event, and it is very important 
to separate these jets properly from the leading jets, which is not an easy task. 
The basic problem is the overlap of two (or more) hadron showers and the 
uncertainties on jet identifications. Under a limiting momenta the
separation can not be done, even more the reconstruction slightly depends on the
applied method, especially with decreasing parton energies.

If momentum exchange become
small, then the background events can not be separated into jets, but considered
together. This is the "underlying event" (UE), where the application of pQCD
is ambiguous and phenomenological descriptions appears to catch certain
features of the hadron ensamble. The underlying event is important especially
in such a cases, when large hadron multiplicities appear in the final state, 
leaving open the question if these extra hadrons are produced during
jet fragmentation or in the center of the collisions with small momentum excange.
In this latter case we expect increasing hadron multiplicities with increasing energy even
in $pp$ collisions, which indicates the enhancement of
entropy production connected to low energy parton-parton interactions.  
This hadron ensamble could display collective behaviours including 
thermalization or the appearance of different flow phenomena.
The characteristics of the UE event in $pp$ collisions
can be connected to the study of heavy ion collisions, where the UE state
is dominantly collective and jet-matter effects can be investigated.
Our aim is to collect information about the jets and UE event, and to study
their measurable features if jet-UE interaction appears.

Underlying event was originally defined by the CDF 
Collaboration~\cite{CDFUE} at the energy of TEVATRON. Since
multiple jet events were very rare, then UE has denoted the remaining hadrons
of a proton-antiproton ($p{\bar p}$) collisions, after leading jets were identified 
and hadrons of jet cones were removed from the whole event. 
The CDF definition of the underlying event is a simple tool.
However, with this definition we are unable to analyze multiple jet 
production.
We have increased the quality of the extracted information by
a modified definition introducing multiple surrounding belts
(SB) around identified jets~\cite{Agocs:2009,Agocs:2010}. 
This new definition immediately leads to a more sophisticated 
analysis of UE.

We have already performed analysis with the new SB-method
and summarized our results in Ref.~\cite{bggmex:2010}. However,
many question remained open. Especially the discussion on quantitative 
differences between the consequences of CDF-definition and 
the SB-method is missing.

Here we focus on this comparision. 
We recall the original CDF-based and our 
SB-based definition of the UE and compare them qualitatively.
We perform quantitative calculations and compare several
physical quantities: (i) the average hadron multiplicities 
within the defined areas relative to the total event multiplicities; 
(ii)  simple geometrical properties such as azimuthal angle distributions 
and pseudorapidity distributions in the close-to-midrapidity regions; 
(iii) transverse momentum distributions in the discussed regions. 

We are interested in these physical quantities, because we want to use
them to look jet-matter interactions and other collective effects 
in nucleus-nucleus ($AA$) collisions. Such investigation demands 
better understanding of UE events in $pp$ collisions, since UE 
can serve as baseline for the detailed study of $AA$ results.

We have performed our analysis on a data sample generated by
PYTHIA 6.2~\cite{pythia62} for $pp$ collisions at 7 TeV. 
The jets have been identified by the cone-based UA1 jet
finder algorithm~\cite{UA1}, setting jet cone angle $R=0.4$.

\section{Generalized definition of the UE}

The basic definition of underlying events is very simple: remove all particles
connected to identified jets from the total event and the wanted UE consists
of the remaining particles.
This kind of definition may have some dependence on the applied
jet-finding algorithm (and its parameters), especially on the
momentum threshold of identified jets. However, if we trust in 
the correctness of recent state-of-the-art jet-finding algorithms  
(see Refs.~\cite{Salam:2009,Salam:2010}), then the remaining 
particles must be unambiguously defined and we can focus on the
properties of this particle ensamble.

\begin{figure}[h]
\resizebox{1.0\columnwidth}{!}{%
 \includegraphics{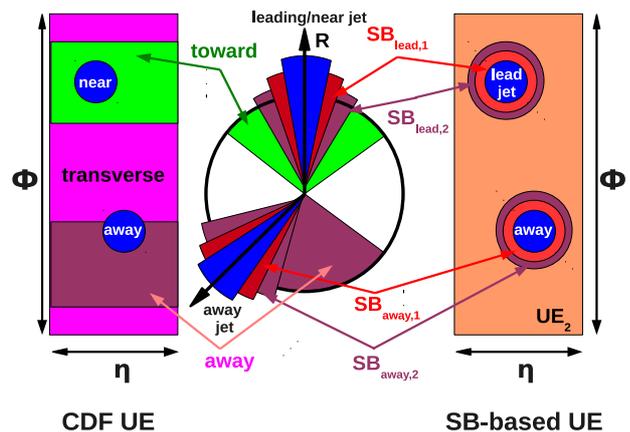} }
  \caption{{\it (Color online)} The schematic view of the underlying event (UE) 
defined by the CDF-method ({\sl left panel}) and the SB-method with
surrounding belts ({\sl right panel}). Detailed
explanation can be found in the text and Ref.~\cite{bggmex:2010}. }
  \label{fig:cdf-sb-ue}
\end{figure}

Historically the CDF collaboration was the first to establish a 
plausible definition to determine UE~\cite{CDFUE}, which  
was commonly used for decades in various analysis. 
This CDF-definition corresponds to jet identification 
in one-jet events, where the second jet is assumed to move automatically
into the away side. The CDF event geometry can be fixed easily, since
the position of the leading jet defines the {\sl toward region},  
and the {\sl away region} will be chosen respectively. 
The left side of Fig.~\ref{fig:cdf-sb-ue} displays this definition.

Our concept was to improve the CDF-based definition on the
following points:

\begin{itemize}
\item to develop an new UE definition, which is strongly connected 
to the identified jets (excluding all hadrons from all identified jets), 
independently on the number of jets.
Thus it can be used for multijet events, also.
\item to gain the capability of investigating the surrounding areas around 
identified jets, without major changes in jet-finding 
parameters. Thus we can study  $pp$ and $AA$ collisions in the same frame
and compare the obtained results, directly.
\end{itemize}

These requirements led us to the definition of surrounding belts (SB) around jet
cones. Thus the UE is defined via excluding hadrons both from jet cones and SBs. 
We are using jet-finding algorithms to define jets and jet-cones, 
than the concentrical surrounding belts are selected by fixing the width of the
belts. The correctness of jet finding algorithm can be easily checked by
comparing the properties of hadrons inside jet cone and surrounding belts.

Fig.~\ref{fig:cdf-sb-ue} serves for the visual comparison of the two
definitions. The CDF-based definition is seen on the {\sl left side},
the SB-based new definition is indicated on the {\sl right side}.
The two definitions can be summarized by means of  
the azimuthal angle, $\Phi$, and (pseudo)rapidity, ($\eta$)  $y$ plane: 

\begin{description}

\item[CDF-based definition] of the UE is based on the subtraction of 
two areas of the whole measured acceptance: one around the identified near jet 
({\sl toward region}) and another to the opposite ({\sl away}) direction. Both 
regions are $\Delta \Phi \times \Delta \eta$-slices of the measured acceptance around 
the near jet and to the opposite, with the full $\Delta \eta $ range and 
limited $\Phi$-range, namely $\Delta \Phi = \pm 60^o$ in azimuth.  

\item[SB-based definition] means the substraction of all hadrons
connected to identified jets and their SB-areas.
Each jet is characterized by an approximate dial-like area, 
around which concentric bands (or rings) are defined. If the 
jet cone angle, $R=\sqrt{\Delta \Phi^2 + \Delta \eta^2}$ is given, then a first 
'$SB_1$' and a second '$SB_2$' surrounding belt will be defined 
for all jets, where the thickness of $\delta R_{SB1}$ and $\delta R_{SB2}$
are introduced. Usually $\delta R_{SB1}=\delta R_{SB2}=0.1$ 
at jet-cone values $R= [0.5,1]$. 
It is easy to see that this UE definition is independent from the
identified jets, even if we have more jets in one collision.

Furthermore, increasing the $\delta R_{SBi}$ values, similar (but not the same) area 
can be covered by the SB-method as by the original CDF-definition. In this way the 
covered areas of the two models 
can become close to be identical at large SB-thickness.
\end{description}


\begin{figure*}[t]
  \centering
  \begin{minipage}[l]{0.48\textwidth}
    \includegraphics[width=\linewidth]{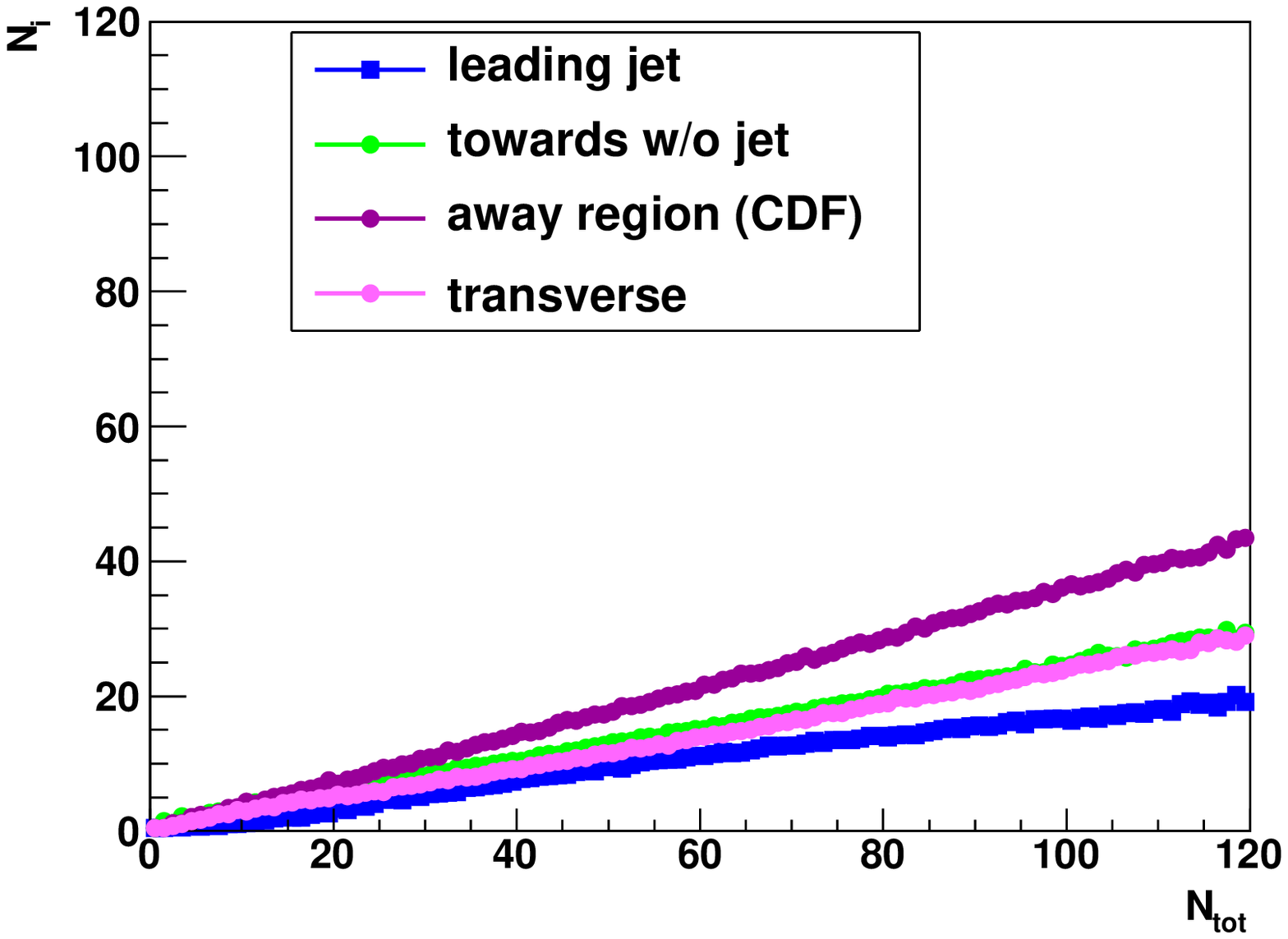}
  \end{minipage}
  \begin{minipage}[r]{0.48\textwidth}
    \includegraphics[width=\linewidth]{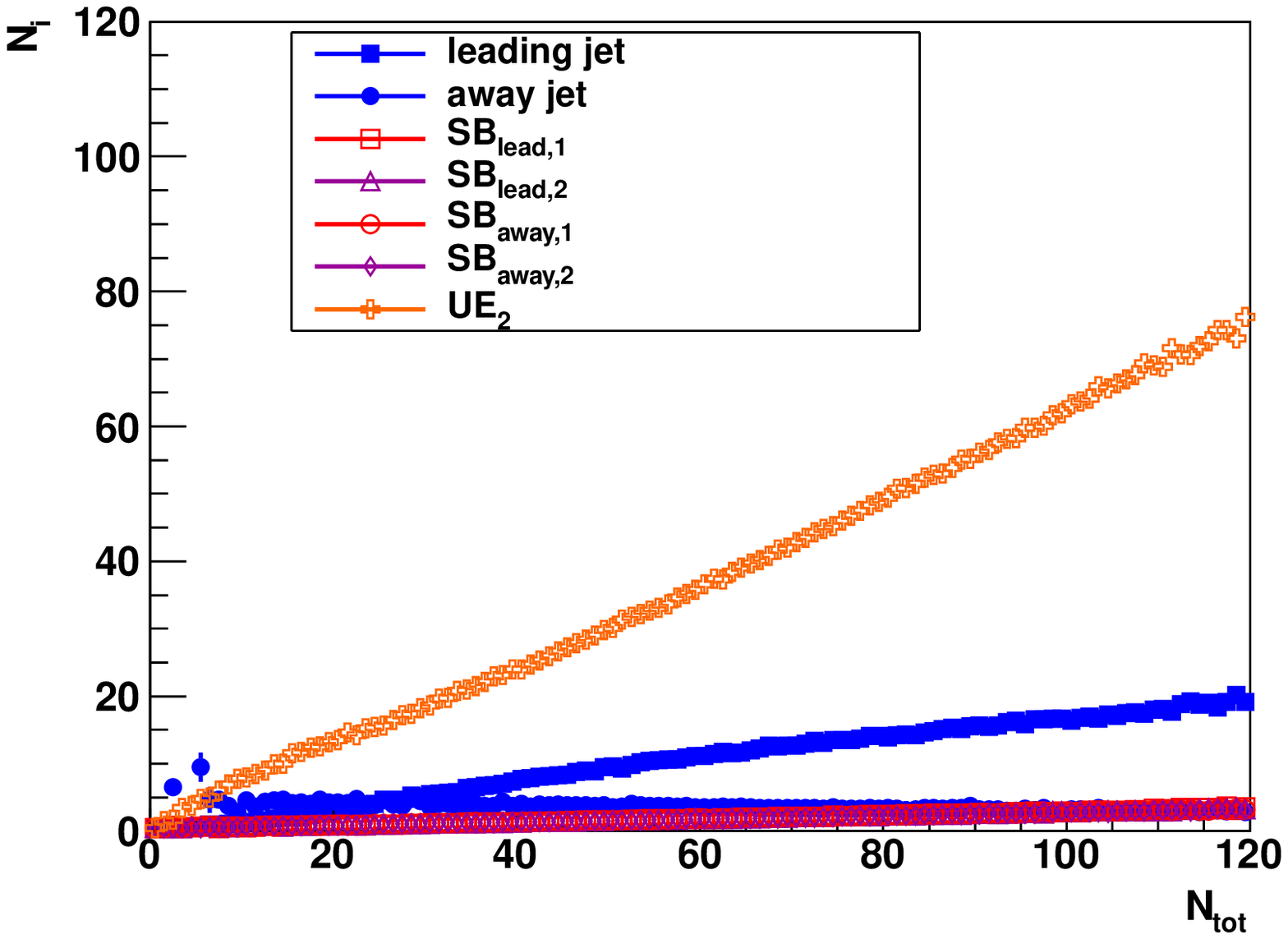}
  \end{minipage}
  \caption{({\it Color online.}) The charged hadron multiplicities, $N_i$, of the selected areas 
depending on the total multiplicities of the events, $N_{tot}$. The CDF-based 
results are displayed on the left panel, SB-based results can be found on the right
panel. Details can be found in the text.}
  \label{fig:n-vs-ntot}
\end{figure*}

\section{Qualitative comparison of different UE sets}

We have already performed an SB-based analysis of a simulated data set for
$pp$ collisions at 7 TeV and published the results in Ref.~\cite{bggmex:2010}.
Here we extend this early investigation for a larger data set of 500 000 $pp$ events 
created by PYTHIA-6 simulation~\cite{pythia62}, applying the Perugia 
tune~\cite{Perugia0}. This sample is similar to the LHC10e14 sample
created at the ALICE experiment for 7 TeV $pp$ collisions. 
In the available sample the jets are identified by the UA1 method~\cite{UA1}. We restrict
our analysis to a limited sample, where the cuts of 
$p_{THardMin} = 10$ GeV/c and $p_{THardMax} = 20$ GeV/c have been applied. 
After these cuts we still had around 156 000 events, which contains at least one jet.
All quantitative results displayed in this paper are extracted from this data sample.

\begin{figure*}[t]
  \centering
  \begin{minipage}[l]{0.48\textwidth}
    \includegraphics[width=\linewidth]{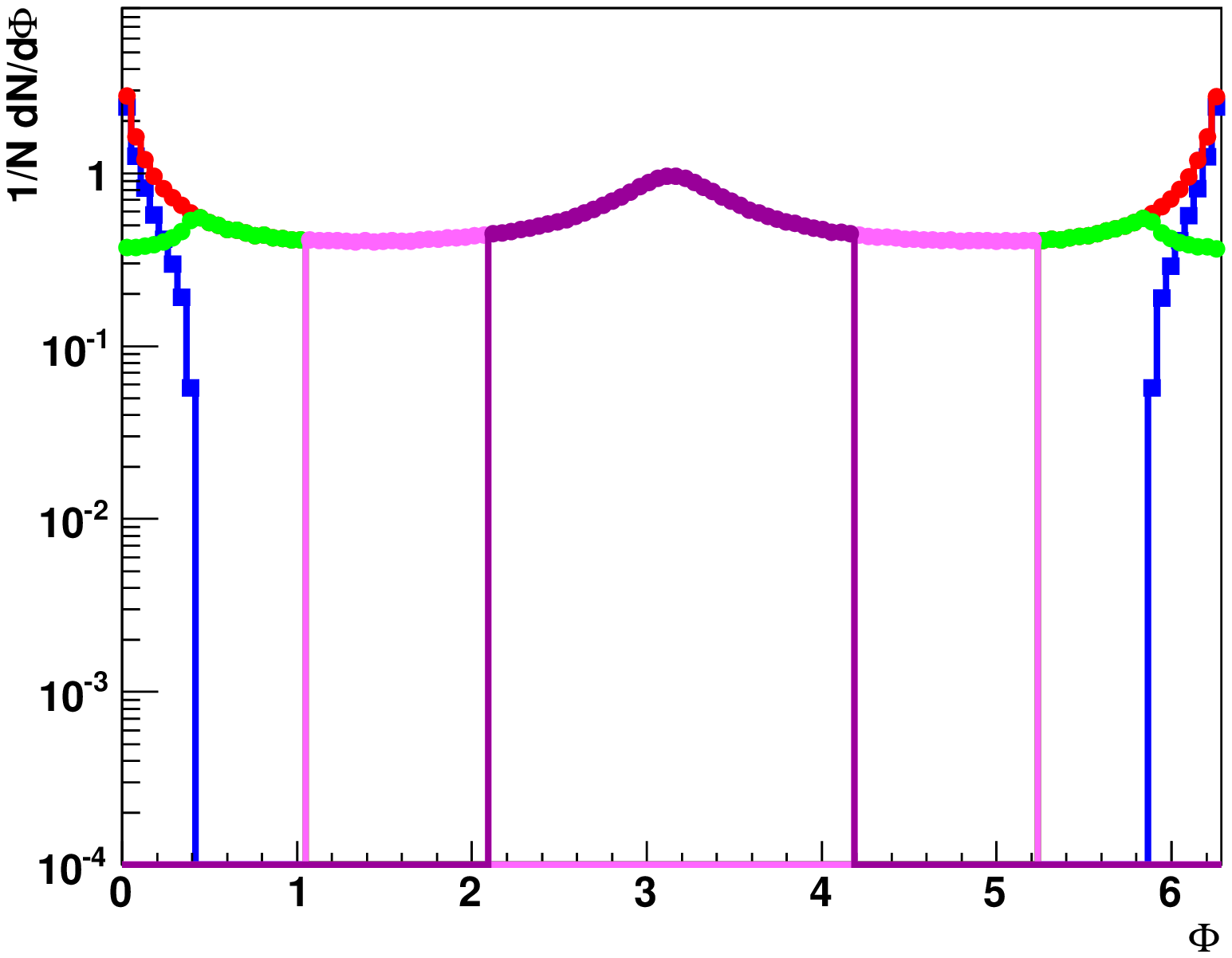}
  \end{minipage}
  \begin{minipage}[r]{0.48\textwidth}
    \includegraphics[width=\linewidth]{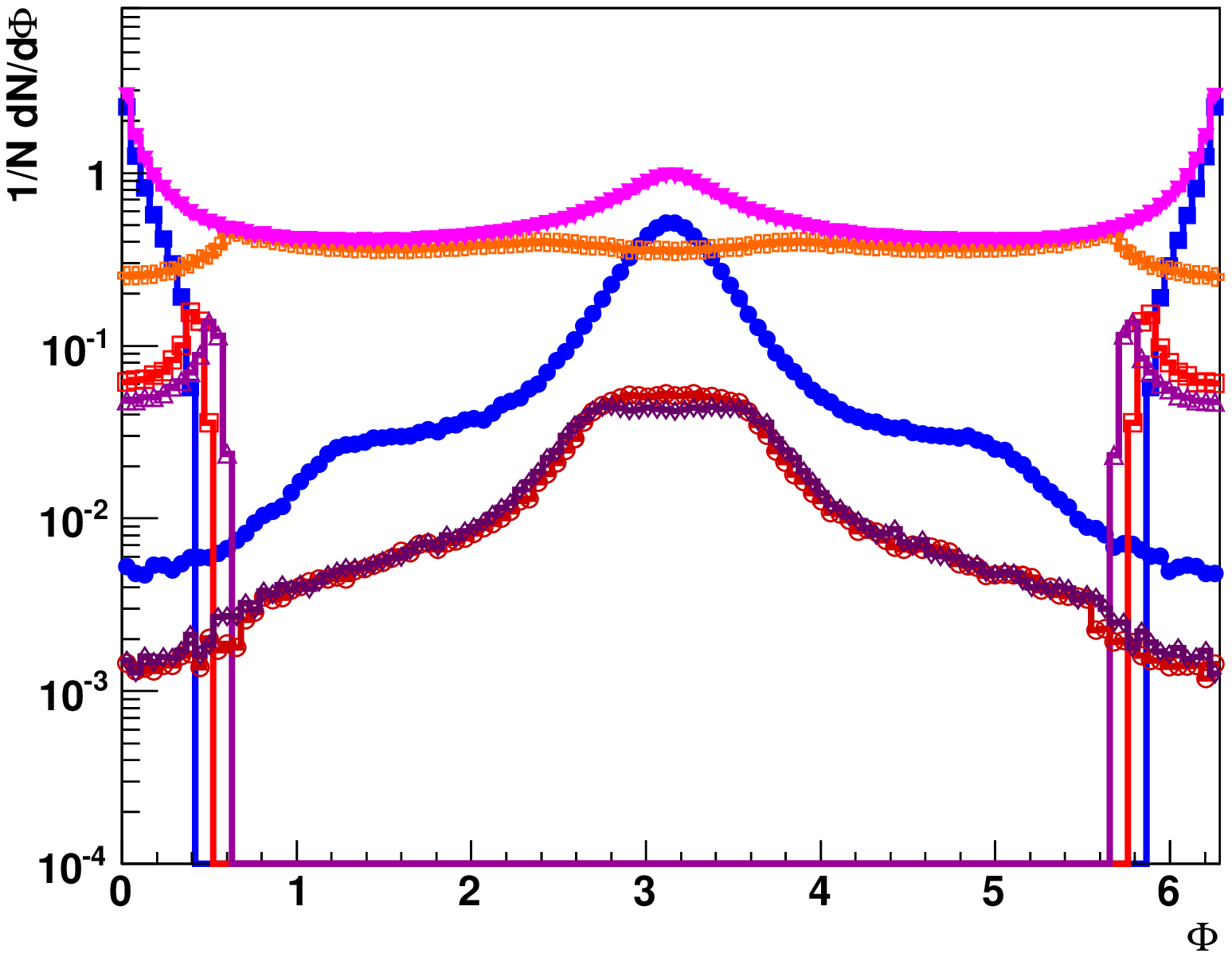}
  \end{minipage}
  \caption{({\it Color online.}) The azimuthal angle ($\Phi$) dependence of particle production 
in the CDF-based ({\sl left panel}) and SB-based ({\sl right panel}) analysis of UE.
Details can be found in the text.}
  \label{fig:phi-dist1}
\end{figure*}

\begin{figure*}[th]
  \centering
  \begin{minipage}[l]{0.48\textwidth}
    \includegraphics[width=\linewidth]{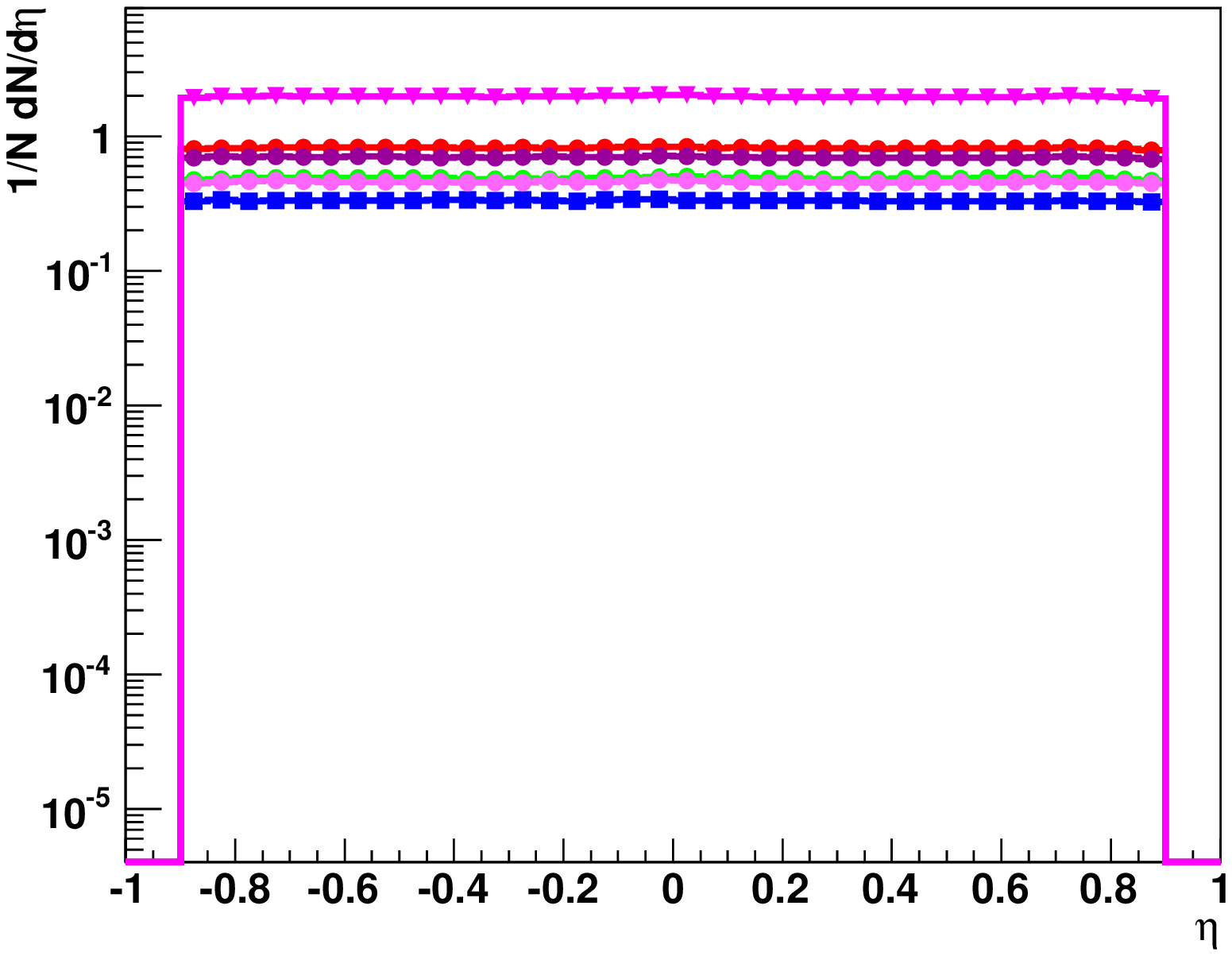}
  \end{minipage}
  \begin{minipage}[r]{0.48\textwidth}
    \includegraphics[width=\linewidth]{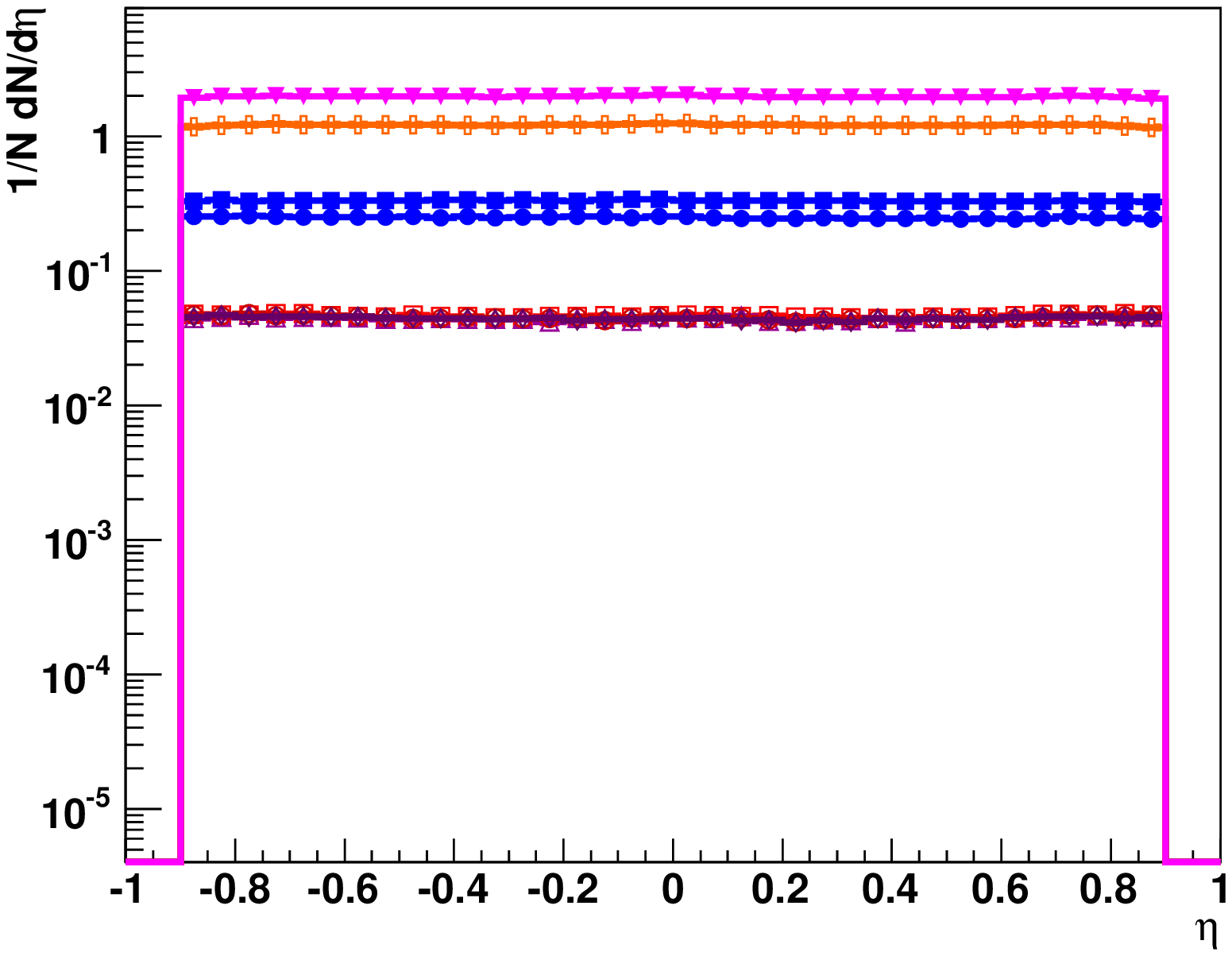}
  \end{minipage}
  \caption{({\it Color online.}) The pseudorapidity ($\eta$) dependence   
in the CDF-based ({\sl left panel}) and SB-based ({\sl right panel}) analysis of UE.
Details can be found in the text.}
  \label{fig:phi-dist2}
\end{figure*}

\subsection{Multiplicity dependences in both analysis}

First we analyse the charged hadron multiplicities in the
various regions of both geometrical definitions.
We apply UA1 jet-finding algorithm to identify jets, than
we collect the multiplicities in the different regions
of the CDF-method and the SB-method. Since the statistics is higher than previously, 
then we expect results with higher precision than in Ref.~\cite{bggmex:2010}.

In Fig.~\ref{fig:n-vs-ntot} we display the multiplicities, $N_i$ of the
different areas depending on the total multiplicities of the events. 
In the {\sl left panel} the displayed results are obtained by the CDF-method.
Here $N_i$ refers to the following contributions:
the multiplicities of the identified 'leading/near jet' 
({\sl blue squares}), the jet-excluded 'toward' area ({\sl green dots}), 
the 'away' side area at the opposite direction ({\sl purple dots}), and the 
CDF-defined UE yield ({\sl pink dots}). 

The {\sl right panel} of Fig.~\ref{fig:n-vs-ntot} displays 
the multiplicities correspond to the
SB-based definition of the underlying event and 
contains detailed information on more areas:  the multiplicities 
of the identified leading jet ({\sl blue squares}), the away side jet 
({\sl blue dots}), multiplicities for the surrounding belts, $SB_{lead,1}$, 
$SB_{lead,2}$, $SB_{away,1}$, and $SB_{away,2}$ ({\sl open red squares, 
open purple triangles, open red circles, open purple diamonds} respectively). 
Finally {\sl orange crosses} denote multiplicities for the newly defined 
underlying events, which collect hadrons outside all identified jets.
We denote this quantity by $UE_2$. (Note, all color codes correspond 
to the areas of Fig~\ref{fig:cdf-sb-ue}.) 

One can see in Fig.~\ref{fig:n-vs-ntot} that all multiplicities, $N_i$, increases
linearly with the total multiplicity in the region $N_{tot}< 120$  
in both cases. Applying the CDF-based definition, the away region gives the 
largest contribution, and the leading jet contribution is the smallest one. 
The transverse area (named as UE) yield an intermediate size contribution.
Moreover,
it is interesting to see, that after excluding the jet from the toward region, 
the remaining area has almost the same multiplicity as the underlying event. This 
shows the correctness of the jet finding algorithm and the "safety" 
of the CDF-based underlying event definition (e.g. 1/3 of the whole acceptance 
is far from any jet-contaminated areas). 

The multiplicities with the SB-based definition differ from the results of the 
CDF-based analysis. The near side jet and away side jet have similar contribution,
since jet-finding algorithm was working properly. The contributions from the belts,  $SB_i$, 
have small multiplicities, since they cover very small areas.
On the other hand, the contribution from the newly defined underlying event, 
$UE_2$ dominates the plot, since it covers almost the whole acceptance. 

In general, the multiplicity fraction of the defined 
areas are almost proportional to the geometrical surface, 
only the jet-content part violates this dependence, 
as Fig.~\ref{fig:n-vs-ntot} displays. Thus, the SB-based $UE_2$ is characterized by 
larger multiplicity value than the CDF-based UE.

\begin{figure*}[t]
  \centering
  \begin{minipage}[l]{0.48\textwidth}
    \includegraphics[width=\linewidth]{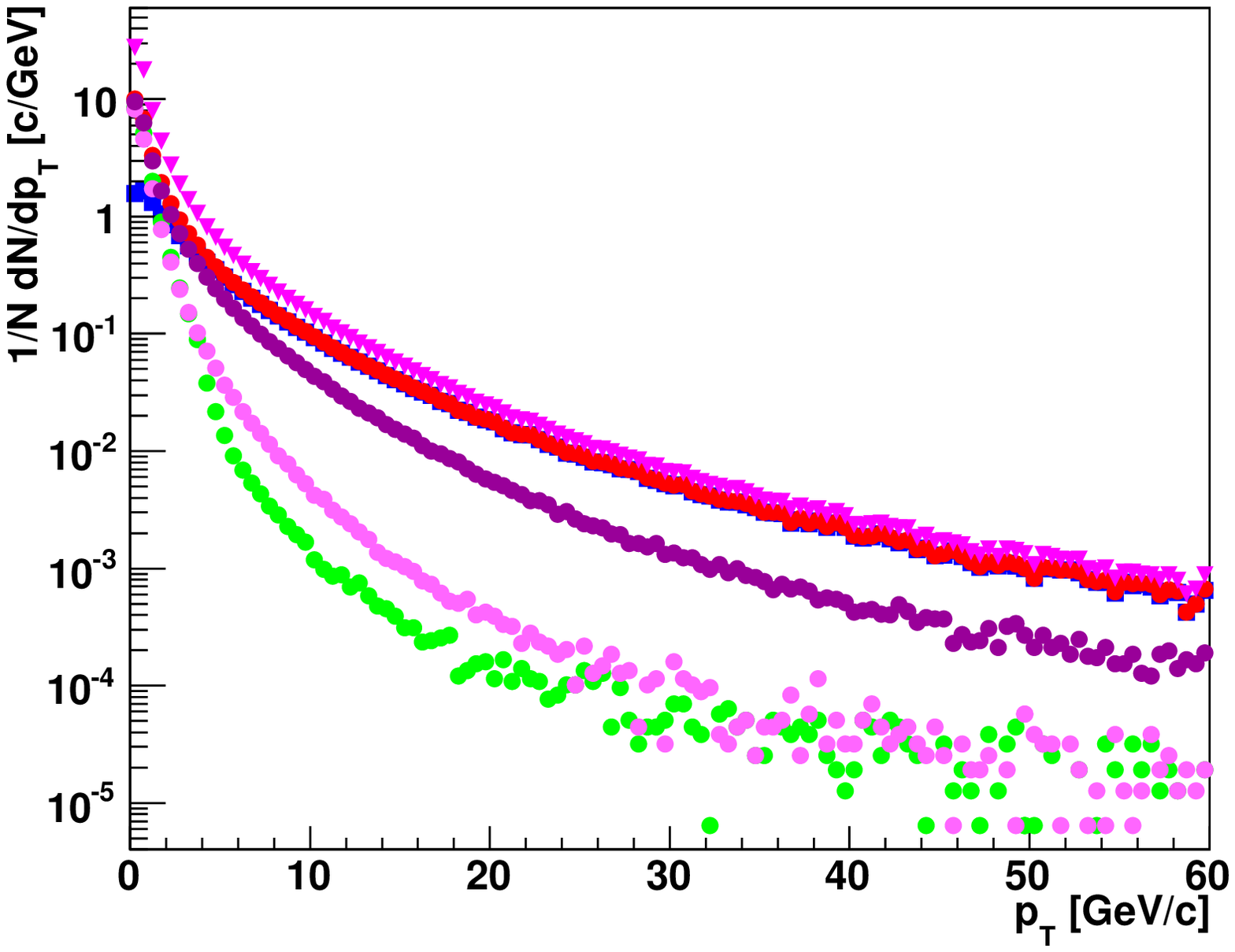}
  \end{minipage}
  \begin{minipage}[r]{0.48\textwidth}
    \includegraphics[width=\linewidth]{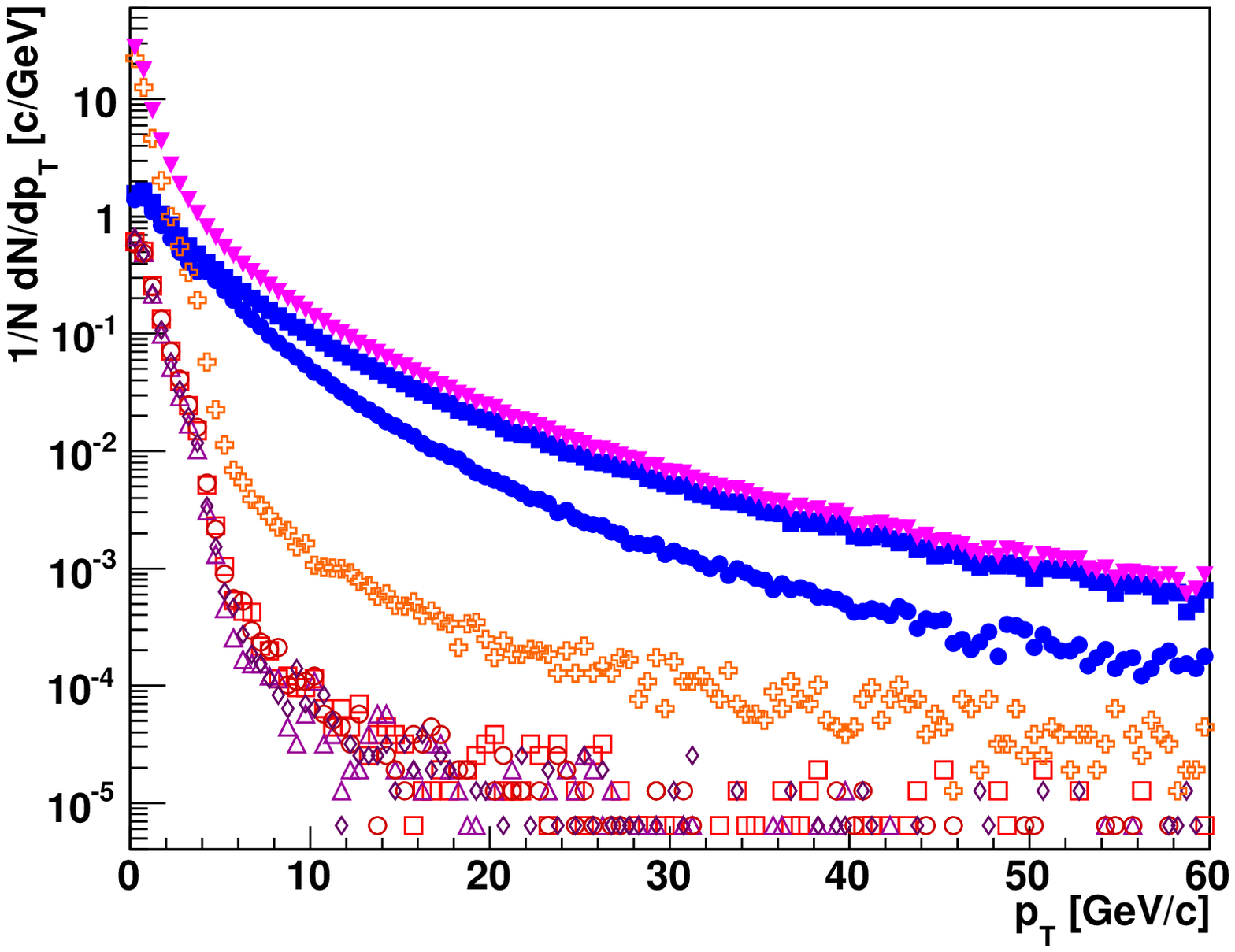}
  \end{minipage}
  \caption{({\it Color online}) The transverse momentum spectra ($p_T$) 
for the CDF-based ({\sl left panel}) and SB-based ({\sl right panel}) underlying 
event and related geometrical regions. More details can be found in the text. }
  \label{fig:phi-dist0}
\end{figure*}

\subsection{Test of geometry}

On Fig.~\ref{fig:phi-dist1} we plotted the azimuth-angle ($\Phi$) 
distribution of the selected areas for CDF-based ({\sl left panel}) and SB-based 
({\sl right panel}) definitions. Here we used the color codes of
Fig.~\ref{fig:cdf-sb-ue} 
and Fig.~\ref{fig:n-vs-ntot}. 

Due to the geometrical roots of the underlying event definitions, the 
azimuth-angle distribution is clearly separated for the CDF-based 
definition but very complicated in the SB-based case. The CDF-based 
results display a clear envelope curve including {\sl red "wings"}. 
The UE yield --- indicated by {\sl magenta} --- is almost flat. 
Opposite to the leading jet, at $\Phi \approx \pi$, the away side 
region is characterized by a well-defined Gaussian-like distribution. 

One can see that the SB-based analysis is more sophisticated,
it carries more information. The jet "wings" at the 
sides and the envelop curve is the same as the results from the
 CDF-based analysis. However, the yields from the 
away-side jet and surrounding belts $SB_{away,1}$ and $SB_{away,2}$ shows 
complex structure ({\sl blue dots, purple diamonds and orange circles}).
The validity of our underlying definition ($UE_2$)is supported by 
the appearance of a wide flat area in $\eta$ (indicated by {\sl orange crosses}),
clearly spearated from jet cone and SB contributions.

On Fig.~\ref{fig:phi-dist1} the pseudorapidity distributions are shown
for CDF-based  ({\sl left side}) 
and SB-based ({\sl right side}) analysis. Since  
we have limited acceptance in the rapidity direction, we focus on the 
close-to-midrapidity regions: $\eta \in \left[ -0.9; 0.9 \right] $. Within 
this areas, both underlying event definitions yield flat rapidity distributions. 
Since particle yields correspond to the surface of the defined areas,
thus $SB$s yields are small. However, the  $UE$ and $UE_2$ contributions 
\ dominate the yields.

\subsection{The $p_T$-distribution for the selected areas}

We are interested in the transverse momentum spectra of charged hadrons detected
in the different areas of the CDF- and SB-setup.
Fig.~\ref{fig:phi-dist0} displays results for $p_T \leq 60$ GeV/c.
We can see relatively large difference between 
the CDF and SB cases.   The {\sl left panel} shows the CDF-based results,
the color encoding corresponds to Fig.~\ref{fig:cdf-sb-ue}. 
The obtained spectra are quite similar for the different regions, 
a close to linear shift characterizes the differences, which is connected
to the proper geometrical surface.
Thus jet-excluded near and the UE areas ({\sl pink and green dots}) display the
smallest yield at high-$p_T$. 
The spectra of away side hadrons ({\sl purple dots}) dominates the yields, 
but without any structure. It seems to us, the CDF-based 
underlying event definition generates separation between  
the spectra of the selected areas. 

On the other hand, the SB-based definition results in a better separation for
the different momentum distributions. The transverse momentum spectra 
within the identified jet cone, plotted by {\sl blue squares and 
points} for comparision and corresponds to the CDF-based definition. 
The underlying event 
has similar spectra, however the SB-based definition results in a more steeper 
distribution at lower-$p_T$ values. Spectra for the surrounding belts
have the lowest yields and starts together with the SB-defined 
underlying event spectra, but with a cut-off at $p_T \approx 10$ GeV/c.

\section{Conclusions}

We have studied underlying event definitions in 
proton-proton collisions at $\sqrt{s} = 7$ TeV. 
We have investigated and compared the multiplicities, azimuth-angle and pseudorapidity 
distributions, and transverse momentum spectra for the CDF-based and  
SB-defined regions.

We tested the multiplicity fraction of the defined regions and repeated our
earlier study~\cite{bggmex:2010}. Using a larger data sample  
a clear separation of the undelying event 
has been found. Comparing the CDF-based and SB-based underlying event definitions
the latter gives a better separation in sense of highest multiplicity relatively 
to the total multiplicits of the event, $N_{tot}$. This originates from the 
more sophysticated definition and the larger geometrical surface to be considered. 
Oppossite to this
the original CDF-based definition does not give a good multplicity separation, 
since away, transverse and jet-excluded toward regions have similar multiplicity 
content.

A more clear explanation of the geometrical distribution is arising from the 
azimuth angle and pseudorapidity distribution. The CDF-based azimuth 
angle distribution is quite clear, it results in two small, almost flat $\Phi$ 
distribution of the UE. In the SB-setup,  
the obtained azimuth angle 
distributions are overlapping, but the whole underlying event region becomes 
a well defined background. Moreover, in the investigated close-to-midrapidity
region, $\left| \eta \right| \lesssim 1 $, the magnitude of the 
constant $\eta$ distrubution is proportional to the area of the CDF- or 
SB-defined regions.

Finally, we compared the transverse momentum spectra 
for the different regions. The SB-method gives a more sophisticated separation
of the charged hadron yields from different regions and its general use is much more
supported to study the properties of the UE and any jet-matter interactions
inside the surrounding belts.

\section*{Acknowledgments}
This work was supported by Hungarian OTKA NK77816, PD73596 and
E\"otv\"os University. One of the authors (GGB) thanks for the J\'anos Bolyai 
Research Scholarship of the Hungarian Academy of Sciences.


\end{document}